# Exploiting PendingIntent Provenance Confusion to Spoof Android SDK Authentication

Ramanpreet Singh Khinda
ramanpreetSinghKhinda@gmail.com
LinkedIn
Sunnyvale, California, USA

**Abstract**

A single authentication bypass in a partner SDK grants attackers the identity of every partner in the ecosystem—and millions of apps use SDKs with exactly this vulnerability. OWASP's 2024 Mobile Top 10 ranks Inadequate Supply Chain Security as the second most critical mobile risk, explicitly identifying third-party SDKs as a primary attack vector. Cross-app mobile SDKs—where a partner application communicates with a platform provider's application via inter-process communication (IPC)—mediate sensitive operations such as content publishing, payment initiation, and identity federation. Unlike embedded libraries that execute within a single app's process, cross-app SDKs require the provider's service to authenticate the calling application at runtime. A pattern sometimes used for this authentication relies on `PendingIntent.getCreatorPackage()` to verify sender identity. We demonstrate that this mechanism exhibits a fundamental *provenance confusion* vulnerability: a `PendingIntent` reliably identifies who *created* it but cannot attest who *presents* it—and this distinction is fatal for authentication. An attacker app with notification access—a user-granted permission held by millions of installed applications—can steal a legitimate partner's `PendingIntent` via `NotificationListenerService` and replay it to impersonate that partner, bypassing authentication entirely. The attack succeeds against both mutable and immutable `PendingIntents` because immutability protects the token's *contents*, not its *provenance*. We systematically evaluate eight Android IPC and application-layer authentication mechanisms against an SDK-specific threat model and present a defense architecture combining Bound Service IPC with kernel-level caller verification via `Binder.getCallingUid()`, supplemented by server-side certificate-hash validation. We validate both the attack and defense through a proof-of-concept implementation on Android 14. Our three-layer scheme provides authentication guarantees rooted in the Linux kernel while remaining scalable across partner ecosystems—new partners are onboarded server-side without SDK updates.

## 1 Introduction

OWASP's 2024 Mobile Top 10—the first update since 2016—ranks *Inadequate Supply Chain Security* as the **second most critical** mobile application risk, explicitly identifying "third-party software libraries, SDKs, vendors" as primary threat vectors that enable attackers to "insert malicious code," "introduce backdoors," or "gain access to backend servers" [40]. This ranking reflects a broader crisis in software supply chains: Sonatype's 2024 State of the Software Supply Chain report documents over 704,000 malicious open-source packages discovered since 2019—a 156% year-over-year increase—while noting that 80% of application dependencies remain un-upgraded for over a year despite 95% having fixes available [47].

Mobile applications sit at the center of this crisis. A typical Android app integrates 15–20 third-party libraries [42], and over 93% of mobile apps use at least one third-party SDK [41]. Third-party libraries have become the largest attack surface for Android applications [54]. The consequences of SDK compromise are ecosystem-wide: in 2020, a single bug in Facebook's iOS SDK caused mass outages twice within 2.5 months, crashing Spotify, TikTok, Pinterest, Tinder, Venmo, and dozens of other major apps—including Apple's own App Store [51]. The Goldoson malware, discovered in 2023, infiltrated 60 legitimate apps with over 100 million combined downloads by hiding inside a third-party advertising library [34]. The Necro Trojan affected up to 11 million Android users through a compromised ad-integration SDK [30]. These incidents demonstrate that SDK vulnerabilities propagate at ecosystem scale—a single compromised SDK becomes a backdoor into every app that embeds it. While much of this concern focuses on embedded libraries, a distinct threat arises in *cross-app* SDKs—where a partner application invokes operations in a platform provider's application via IPC. Authentication failures at this boundary enable partner impersonation with ecosystem-wide consequences.

In cross-app integration, a platform provider distributes an SDK that partner applications embed to communicate with the provider's own application via IPC. This model—used for content publishing, payment authorization, and identity federation—presents a distinctive security challenge: the platform provider's service must authenticate which partner application is invoking it, but all apps—legitimate partners and potential attackers—execute on the same device. Consider a content distribution platform ("Amplifier") that provides an SDK enabling partner apps to publish content on its network. When a partner app ("MyBeats") invokes the SDK's publish API, Amplifier must verify that the request genuinely originates from MyBeats and not from a malicious app masquerading as a legitimate partner. Backend API keys are insufficient—they are embedded in the client binary and trivially extractable via reverse engineering [61]. Server-side credentials alone (e.g., per-partner OAuth tokens) cannot distinguish a legitimate partner's request from an attacker's, since both arrive through the same SDK code path on the device. The platform therefore requires a *runtime* mechanism to verify the calling app's identity at the IPC boundary.

One solution uses `PendingIntent` tokens for this purpose. The SDK library (embedded in the partner app) creates a `PendingIntent` representing the partner's identity and includes it with each request to the SDK provider's service; the service extracts the creator's package name via `getCreatorPackage()` and checks it against a partner allowlist. While Android's security documentation explicitly warns against this pattern due to exploitation risks [11], the method is actively used in production. WhatsApp's one-tap autofill authentication flow uses `getCreatorPackage()` to verify partner app identity—a pattern that propagates through business messaging SDKs (Alibaba Cloud, Webex Connect, and others document identical logic [3, 55]) and that Meta is formally deprecating in April 2026 in favor of a handshake-ID model [35]. Chromium's IntentHandler checks `getCreatorPackage()` for token validation [18]. The Termux plugin architecture uses it to verify official app origin [50]. Even Android's own `MediaButtonReceiverHolder` uses `getCreatorPackage()` to filter media button receivers by creator package [14]. The intuitive appeal of "check the creator" reasoning persists, and the mechanism appears sound: a PendingIntent's creator identity is set at construction time and cannot be altered.

We show that this reasoning contains a critical gap. A `PendingIntent` identifies who *created* it but provides no guarantee about who *presents* it. We term this disconnect *provenance confusion*: the conflation of token origin with token presenter. Any app that obtains a reference to another app's `PendingIntent`—for example, by extracting it from that app's notifications via `NotificationListenerService`—can present it to the SDK provider's service and authenticate as the original creator. Since the vast majority of Android users receive push notifications and every tappable notification must contain a `PendingIntent` [9], the attack surface is vast.

While Android's security documentation warns about PendingIntent theft in general [11], neither the platform guidance nor prior academic work addresses the *SDK partner authentication* context. In privilege escalation attacks [56], the victim is an *end user* whose permissions are hijacked. In our attack, the victim is a *business partner* whose brand identity and platform quota are hijacked—the attacker publishes content, initiates payments, or federates identity under the partner's name.

**Contributions.** We make four contributions:

(1) **Attack.** We present a practical identity spoofing attack against PendingIntent-based SDK authentication. The attack exploits `NotificationListenerService` to steal a legitimate partner's `PendingIntent` and replay it, requiring no platform exploits and succeeding against both mutable and immutable tokens (§3).

(2) **Systematic analysis.** We evaluate eight authentication mechanisms—`startActivityForResult`, `getReferrer()`, `PendingIntent.getCreatorPackage()`, BroadcastReceiver permissions, ContentProvider permissions, signature/knownSigners permissions, PKCE, and Bound Services—against an SDK-specific threat model (§4).

(3) **Defense.** We present a three-layer defense architecture combining Bound Service IPC with kernel-level Binder UID verification and server-side certificate-hash validation, providing authentication guarantees that are unspoofable from unprivileged application context (§5).

(4) **Evaluation.** We validate both the attack and defense through a proof-of-concept implementation, demonstrating 100% attack success against vulnerable SDKs and 100% defense effectiveness (§6).

## 2 Background and Threat Model

### 2.1 Android IPC and PendingIntent

Android apps run in isolated sandboxes, each assigned a unique Linux UID at install time. Inter-process communication (IPC) occurs through the Binder kernel driver, which mediates all cross-process calls and provides kernel-enforced caller identity [16]. A `PendingIntent` encapsulates an `Intent` together with its creator's identity and permissions [9]. When executed, the system performs the action with the *creator's* identity, regardless of which process triggered execution.

Every tappable notification must include a `PendingIntent` [9]. Since Android 12 (API 31), all PendingIntents must declare mutability via `FLAG_IMMUTABLE` or `FLAG_MUTABLE`. Crucially, *immutability constrains the token's contents but does not restrict who may hold or present it*. As of late 2025, approximately 90% of active Android devices run Android 12 or higher [12], making the `FLAG_IMMUTABLE` requirement practically universal.

The security implications of `PendingIntent` misuse are well documented. PITracker identified 72 PendingIntent-related CVEs in 2020–2021 alone and found 2,939 vulnerable `PendingIntent` objects across AOSP [56]. High-severity vulnerabilities have affected both AOSP (CVE-2020-0188, CVE-2020-0294) and vendor components, with Samsung alone disclosing four PendingIntent hijacking CVEs between 2021 and 2023 [44]. These vulnerabilities predominantly involve privilege escalation and data leakage; our work identifies a distinct attack class—authentication bypass in SDK ecosystems.

### 2.2 NotificationListenerService

`NotificationListenerService` (NLS) is an Android framework component that grants registered apps access to all posted notifications on a device [8]. An NLS-enabled app receives callbacks for every notification posted by any application, including full access to the notification's `PendingIntent` objects (content intent, action intents, delete intent).

NLS access requires explicit user opt-in through a system settings dialog—it is *not* a system-level permission but rather a user-granted capability similar to accessibility services. No root access, ADB commands, or platform exploits are required. Despite this friction, NLS is widely granted: legitimate use cases include notification managers, wearable companion apps (Samsung Galaxy Wearable, Fitbit), automation tools (Tasker, IFTTT), digital wellbeing apps, and accessibility tools for visually impaired users [32]. MITRE ATT&CK documents NLS abuse as technique T1517 ("Access Notifications"), noting that malicious apps use it to intercept OTPs, banking alerts, and messaging content [37].

Three active malware families have weaponized NLS for credential theft: Tria Stealer produced over 107,000 samples across 2.5 years of operation [31]; Aberebot targets banking apps by intercepting 2FA codes [20]; and OtpSteal specifically harvests one-time passwords from notification content [52]. The S.O.V.A. banking trojan "abuses [notification access] to read and modify notifications received on the device," enabling interception of OTPs and personal messages [19]. These campaigns confirm that NLS is a well-established tool in the arsenal of Android banking and surveillance malware.

Android 15 introduced a new `RECEIVE_SENSITIVE_NOTIFICATIONS` permission (`signature|role` protection) and OTP-specific content masking for NLS apps [5]. However, these mitigations target data exfiltration—they mask *notification content* (OTP codes) but do *not* restrict access to the `PendingIntent` objects embedded in notifications. An NLS-enabled attacker on Android 15 can still extract PendingIntents from every notification—only the text content is masked. Android 15 also extends the Restricted Settings framework to block *sideloaded* apps from obtaining NLS access, but this restriction does not apply to apps distributed through the Play Store or compliant third-party stores—a malicious app that passes store review (as documented in prior work [30, 34]) retains full NLS access. Our attack vector remains fully functional on Android 15.

## 2.3 SDK Partner Authentication

In a partner SDK ecosystem, a platform distributes an SDK library that third-party partner apps embed. The SDK mediates access to platform capabilities (e.g., content publishing, payment processing). The platform must ensure that only registered partner apps invoke these capabilities.

The scale of SDK ecosystems amplifies the consequences of authentication failures. Partner SDKs are distributed to partner apps via public or private repositories (AARs/Maven artifacts). A typical Android app integrates 15–20 third-party SDKs [42], and over 93% of mobile apps use at least one SDK [41]. SDK supply-chain compromises have repeatedly demonstrated ecosystem-wide impact: the SourMint malware (hidden in Mintegral's advertising SDK) affected 1,200+ iOS apps with approximately 300 million downloads per month [46]; the X-Mode location SDK was banned by Apple and Google in 2020 and later prohibited by FTC order from selling sensitive location data [23]; OneAudience and MobiBurn SDKs harvested Facebook and Twitter user data, leading to platform bans and litigation [36].

The PendingIntent-based authentication pattern works as follows: (1) the SDK library creates a `PendingIntent` representing the partner app's identity; (2) the SDK provider's service extracts the creator package via `getCreatorPackage()`; (3) the service checks the creator package against a server-maintained allowlist of registered partners. If the package matches, the request is authenticated.

## 2.4 Threat Model

**Attacker goal.** Impersonate a legitimate partner app to invoke privileged SDK operations (e.g., publish content under the partner's identity, initiate payments, federate identity).

**Attacker capabilities.** The attacker controls a malicious app installed on the same device as the target partner app and the SDK provider's app. The attacker has obtained `NotificationListenerService` access (requires user opt-in via Settings). The attacker does *not* have root access, cannot modify other apps' binaries, and cannot exploit platform-level vulnerabilities.

**Trust assumptions.** The Android kernel and Binder driver are trusted. The package manager's UID-to-package mapping and certificate verification are trusted. The attacker cannot forge Linux UIDs or signing certificates from unprivileged context. These assumptions align with Android's security model and are validated by academic analysis of Binder's security properties [16, 25].

## 3 The Attack

### 3.1 Overview

The attack exploits the provenance confusion inherent in PendingIntent-based authentication: `getCreatorPackage()` returns the identity of the app that *created* the token, not the app that *presents* it. An attacker who obtains a reference to a partner's `PendingIntent` can present it to the SDK provider's service and authenticate as that partner.

Figure 1 contrasts normal authentication with the attack. In the normal flow (left), the SDK library creates a `PendingIntent` on behalf of MyBeats, and the SDK provider's service verifies the creator identity. In the attack (right), a malicious app extracts MyBeats' `PendingIntent` from a notification and replays it to the SDK provider's service, which returns the legitimate partner's identity and authenticates the attacker.

### 3.2 Attack Steps

*Step 1: Credential Harvest.* The attacker app registers a `NotificationListenerService` (NLS). When MyBeats posts any notification (e.g., playback status, download progress), the NLS callback receives the full `StatusBarNotification` object, including all embedded `PendingIntent` references:

**Listing 1: Extracting a PendingIntent from notification.**

```
class StealerNLS : NotificationListenerService() {
  override fun onNotificationPosted(sbn: StatusBarNotification) {
    if (sbn.packageName == "com.mybeats.app") {
      // Extract content PendingIntent
      val stolenPI = sbn.notification.contentIntent
      CredentialCache.store(sbn.packageName, stolenPI)
    }
  }
}
```

*Step 2: Identity Spoofing.* The attacker invokes the SDK's API, substituting the stolen `PendingIntent` for its own:

**Listing 2: Replaying stolen PendingIntent to SDK.**

```
// Attacker's app - includes Amplifier SDK dependency
val stolenPI = CredentialCache.get("com.mybeats.app")
AmplifierSDK.publish(
    content = maliciousPayload,
    authToken = stolenPI // SDK sees creator = MyBeats
)
```

*Step 3: Authentication Bypass.* The SDK provider's service calls `stolenPI.getCreatorPackage()`, which returns `"com.mybeats.app"`—the original creator's identity, not the attacker's. The service matches this against the partner allowlist and authorizes the request:

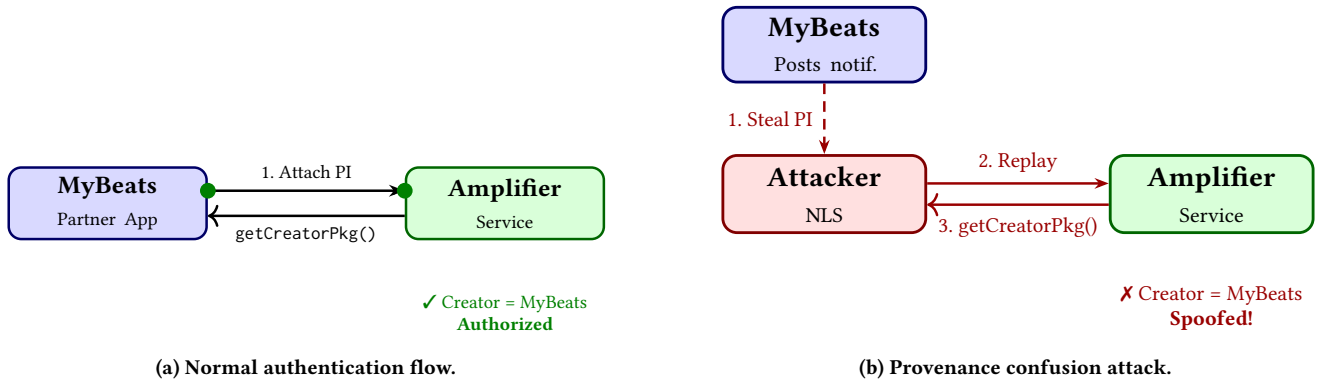

(a) Normal authentication flow.

(b) Provenance confusion attack.

Figure 1: Normal authentication (left) vs. provenance confusion attack (right). In both cases, `getCreatorPackage()` returns `"com.mybeats.app"`. The SDK provider's service cannot distinguish the legitimate partner from the attacker.

**Listing 3: Vulnerable authentication logic in SDK provider's service.**

```
fun authenticate(pi: PendingIntent): Boolean {
  val caller = pi.creatorPackage // Returns "com.mybeats.app"
  return partnerAllowlist.contains(caller) // true!
}
```

## 3.3 Why Immutability Does Not Help

Since Android 12, PendingIntents must declare mutability. One might expect that `FLAG_IMMUTABLE` prevents this attack. It does not: immutability prevents modification of the *encapsulated Intent's extras and action*, but does not prevent a different process from *holding and presenting* the token. The token's creator identity is a read-only property set at construction; the token's *presenter* is never recorded or verified. This is the core of provenance confusion: the API conflates "who made this credential" with "who is authenticating with this credential."

Paradoxically, the introduction of `FLAG_IMMUTABLE` may have *increased* false confidence: developers who mark tokens as immutable may believe they have addressed the security guidance, when immutability is orthogonal to provenance confusion.

## 3.4 Why Content Cannot Be Embedded

One might ask why the SDK does not require content to be embedded within the PendingIntent's immutable extras, binding the authenticated identity to the payload. The answer is architectural: **PendingIntent contents are opaque to the receiver**.

Figure 2 illustrates this constraint. Android provides no public API for extracting the wrapped Intent or its extras from a received PendingIntent—the receiver can only invoke the token via `send()` [9]. There is no `PendingIntent.getIntent()` method. The receiver can check `creatorPackage` and `creatorUid`, but cannot read what is inside the encapsulated Intent.

Consequently, any content the SDK provider's service must *process* (not merely forward) must be passed as a separate, readable parameter. This architectural requirement creates the credential-content separation that enables our attack: the PendingIntent serves only as an identity badge, while the actual payload travels alongside it under the attacker's full control.

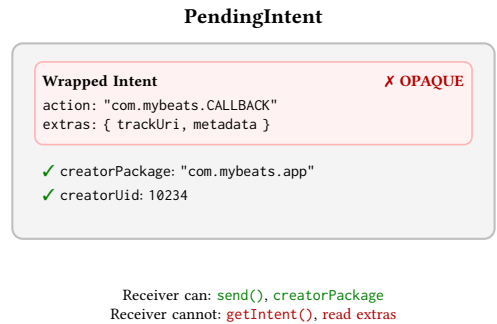

Figure 2: PendingIntent as a black box. The receiver can invoke the token and check creator identity, but cannot extract the wrapped Intent or its extras. Content must therefore travel separately.

## 3.5 Attack Surface Scope

*Attacker's SDK access path.* SDK libraries are distributed to partner apps via public or private repositories (AARs/Maven artifacts). The SDK provider's Bound Service is exported to allow partner binding. The attacker includes the SDK library and calls its API directly—the authentication check in the provider's service is the *only* barrier.

*Structural prevalence.* Every tappable notification requires a PendingIntent object [9]. Users receive 46 notifications per day on average, and 91% opt in to notifications [32]—so dozens of identity-carrying PendingIntent objects are generated and accessible daily. An NLS-enabled attacker can harvest them from *all* notifying apps. Prior work identified 2,939 vulnerable PendingIntent objects across AOSP and 72 PendingIntent-related CVEs in 2020–2021 [56], confirming systemic misuse.

*Attack feasibility.* For the attack to succeed: (1) the target partner app must post notifications (near-universal), (2) the attacker must have NLS access (held by widely-installed utility apps), and (3) the target SDK must use `getCreatorPackage()` for authentication. Given the structural prevalence of conditions (1) and (2), any SDK exhibiting the vulnerable pattern is immediately exploitable at scale.

*Ecosystem-scale consequences.* A vulnerability in a widely-deployed SDK cascades across its entire partner ecosystem. The Goldoson malware compromised 60 apps with 100M+ downloads [34]. The Necro Trojan affected 11M users through compromised ad SDKs [30]. SourMint infiltrated 1,200+ iOS apps with 300M monthly downloads [46]. An authentication bypass in a partner SDK would grant the attacker the identity and privileges of *any* partner in the ecosystem: a single point of compromise with ecosystem-wide impact.

## 4 Analysis of IPC Authentication Mechanisms

We systematically evaluate eight authentication mechanisms for their suitability as SDK partner authentication primitives. Table 1 summarizes the results across five security-relevant properties.

`startActivityForResult` / `getReferrer()`. Both mechanisms rely on the calling Activity's task stack. The referrer is populated by the system but can be spoofed via task hijacking [33]. Neither provides kernel-backed identity. Additionally, both require an Activity context, which SDK calls from background services or ContentProviders cannot provide.

`PendingIntent.getCreatorPackage()`. As demonstrated in §3, this mechanism is vulnerable to provenance confusion. The creator identity is authentic but the presenter is unverified.

**BroadcastReceiver / ContentProvider permissions.** Custom permissions declared in the SDK's manifest can gate access. However, these permissions must be statically declared by each partner app in its manifest at build time, preventing server-side partner onboarding. Additionally, custom permissions are subject to race conditions during installation [27].

**Signature / `knownSigners` permissions.** Android's signature-level permissions restrict access to apps signed with the same certificate—suitable for first-party suites but unusable across partner ecosystems with independent signing keys. The `knownSigners` flag (API 30+) allows enumerating specific partner certificate hashes in the manifest, but each new partner or key rotation requires an SDK host-app update, creating the same scalability bottleneck as custom permissions.

**PKCE (Proof Key for Code Exchange).** PKCE [43] ensures that the entity completing an authorization flow is the same entity that initiated it (session continuity). However, PKCE does not verify *who* the initiator is. An attacker can initiate their own valid PKCE flow while claiming to be a trusted partner; PKCE succeeds because the attacker completes their own flow. PKCE prevents interception of authorization codes mid-flow, but does not prevent impersonation— an attacker who initiates the flow *is* the legitimate initiator of that flow. Our threat model requires authenticating sender identity, which PKCE's challenge-response mechanism does not provide.

**Play Integrity API.** Google's Play Integrity API [10] provides app-to-server attestation: the server can verify that requests originate from a genuine, unmodified app binary. However, Play Integrity does not address *on-device* app-to-app authentication—it attests app identity to a remote server, not to another local process. Additionally, Play Integrity requires Google Play Services (unavailable on some devices and regions), imposes per-call latency (100ms+ network round-trip), and enforces daily quota limits unsuitable for high-frequency SDK calls. Play Integrity complements but does not replace on-device IPC authentication; we exclude it from Table 1 as it addresses a different trust boundary (app-to-server rather than app-to-app).

**Bound Service + `Binder.getCallingUid()`.** A Bound Service receives IPC calls through the Binder kernel driver, which populates the caller's UID from the Linux kernel's process credentials (task_struct → cred → uid). This UID is unforgeable from unprivileged application context—a property validated by academic analysis of 170+ system services across six Android versions [25] and confirmed by formal security models [16]. The mechanism supports server-side scalability (new partners registered without manifest changes), bidirectional communication, and the strongest available identity guarantee on Android.

Only Bound Service IPC satisfies all five properties simultaneously. This makes it the only mechanism suitable as the sole authentication primitive in a partner SDK ecosystem.

## 5 Defense Architecture

We present a three-layer defense that replaces PendingIntent-based authentication with Bound Service IPC, augmented by certificate verification and server-side validation. Figure 3 illustrates the architecture.

A natural question is whether backend API authentication alone suffices. It does not: the SDK executes within the *partner's* process on an untrusted device. Without device-level identity verification, the backend cannot distinguish a legitimate partner's request from an attacker's—both arrive via the same SDK code path with valid API credentials. Caller identity must be verified *on-device* before the request leaves the process.

The combination of device-side and server-side verification must satisfy constraints unique to partner ecosystems: (1) new partners must be onboarded without SDK updates, (2) the mechanism must scale to hundreds of partners, and (3) backward compatibility with deployed partner binaries. These constraints eliminate static-permission approaches and motivate the layered design we present.

### 5.1 Layer 1: Kernel-Level Caller Identity

The SDK provider's app exposes its API through an Android Bound Service. Bound Services do not require foreground execution or user-visible UI—the SDK library (embedded in the partner app) binds to the provider's service, completes the IPC call, and unbinds while the user remains in the partner app throughout with no app switching required. When the SDK library calls `bindService()`, the Binder kernel driver establishes an IPC channel and makes

Table 1: Systematic comparison of Android IPC authentication mechanisms for SDK partner verification. ✓ = satisfies property; ✗ = does not satisfy; ∼ = partially satisfies. "Kernel-backed" indicates the identity guarantee is enforced by the Linux kernel rather than application-layer logic.

| Mechanism | Kernel-backed identity | Unforgeable from app context | Replay-resistant | Scalable (no manifest edit) | Bidirectional |
|---|---|---|---|---|---|
| startActivityForResult | ✗ | ✗ | ✓ | ✓ | ✗ |
| getReferrer() | ✗ | ✗ | ✓ | ✓ | ✗ |
| PI.getCreatorPackage() | ✗ | ✗ | ✗ | ✓ | ✓ |
| BroadcastReceiver (perm.) | ∼ | ✓ | ✓ | ✗ | ✗ |
| ContentProvider (perm.) | ∼ | ✓ | ✓ | ✗ | ✗ |
| Signature/knownSigners | ∼ | ✓ | ✓ | ✗ | ✗ |
| PKCE | ✗ | ✗ | ✓ | ✓ | ✗ |
| **Bound Service + Binder UID** | ✓ | ✓ | ✓ | ✓ | ✓ |

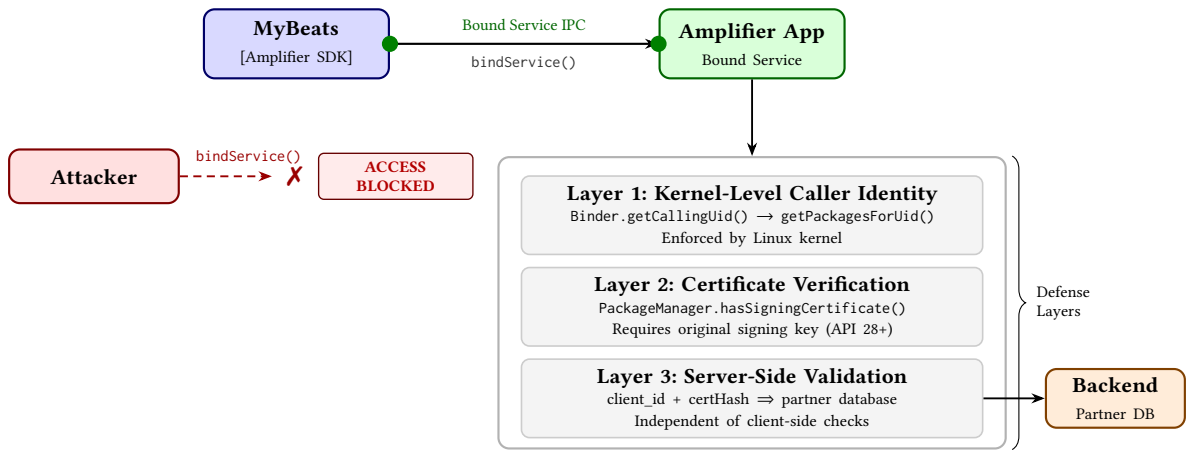

Figure 3: Three-layer defense architecture. The SDK provider replaces PendingIntent-based authentication with a Bound Service IPC channel. Layer 1 obtains the caller's UID from the Binder kernel driver (unspoofable from unprivileged context). Layer 2 verifies the caller's signing certificate against registered hashes. Layer 3 validates credentials server-side, independent of client-side checks. An attacker binding to the service is immediately identified by their true UID and rejected.

the caller's UID available via `Binder.getCallingUid()`. The provider's service resolves this UID to a package name using `PackageManager.getPackagesForUid()`:

**Listing 4: Kernel-backed caller verification in SDK provider's service.**

```
1  override fun onTransact(...): Boolean {
2    val callerUid = Binder.getCallingUid()
3    val callerPkg = packageManager
4      .getPackagesForUid(callerUid)?.firstOrNull()
5      ?: return false  // Unknown caller
6    // Layer 2: verify signing certificate
7    if (!verifyCertificate(callerPkg)) return false
8    // Layer 3: server-side validation
9    return serverValidate(callerPkg, getCertHash(callerPkg))
10 }
```

This UID is derived from the Linux kernel's process credentials and cannot be forged by an unprivileged application. Unlike `getCreatorPackage()`, `getCallingUid()` identifies the *actual caller* of the current IPC transaction—not the creator of a transferable token.

## 5.2 Layer 2: Certificate Verification

After resolving the caller's package name, the provider's service verifies the app's signing certificate against a registered hash using the `hasSigningCertificate()` method in `PackageManager` (API 28+). This layer is essential because package names alone are insufficient: an attacker can sideload a malicious APK with the same package name as a legitimate partner on any device where that partner is not installed.

Sideloading is not a marginal vector. Security reports indicate that sideloading is particularly prevalent in regions where Google Play is restricted or alternative stores are common, including China, India, and parts of Asia-Pacific [60]. Google reports that devices installing apps from unknown sources have significantly higher rates of potentially harmful applications (PHAs) than Play-only devices [24]. Real-world package-name spoofing campaigns have impersonated Telegram and WhatsApp using identical package identifiers distributed through lookalike websites [21]. Certificate verification ensures that even if a sideloaded app matches the partner's package name, the signing certificate will not match, and the request will be rejected.

### 5.3 Layer 3: Server-Side Validation

A natural question is why Layer 2 (certificate verification) requires server-side validation rather than hardcoded certificate hashes. Consider two insufficient alternatives:

*Alternative A: Hardcoded certificate hashes.* The provider's service could embed known partner certificate hashes and verify locally. This is secure—certificates cannot be spoofed—but does not scale. Each new partner requires an update to the SDK provider's app. Worse, when a partner rotates their signing key (required periodically for security hygiene), they are locked out until the SDK releases an update with the new hash. In ecosystems with hundreds of partners and independent release cycles, this creates an untenable coordination burden.

*Alternative B: Server-side validation without certificate hash.* The provider's service could transmit only (packageName, client_id) to the server, omitting the certificate. This fails against sideloading: if the legitimate partner app is not installed on a device, an attacker can sideload a malicious APK with the *same package name* and a stolen client_id. The server sees a valid package name and valid client_id, and the attack succeeds.

*Our approach: Server-side validation with certificate hash.* The provider's service transmits the verified (packageName, certHash) tuple alongside a developer-provisioned client_id. The server validates the full (packageName, certHash, client_id) triple against its registered partner database. This combines the security of certificate verification (defeating sideload attacks) with the scalability of server-side management (new partners onboarded without SDK updates, certificate rotations handled by updating the server database).

The distinction between packageName and client_id enables fine-grained control: packageName is a per-app identity (verified on-device), while client_id is a per-partner identity (verified server-side). A single partner organization may operate multiple apps sharing a common client_id for quota and policy management.

This layer also ensures that even if a device's local state is compromised, the server maintains an independent trust anchor, and enables partner lifecycle management: revoked partners are de-authorized instantly without any SDK or host-app updates.

### 5.4 Security Properties

The security of this scheme does not rely on secrecy of the transmitted values. An attacker may know a legitimate partner's package name, certificate hash, and client_id. The defense operates at two layers. First, the backend only accepts requests from the SDK provider's application, enforced via certificate pinning or Play Integrity attestation—preventing attackers from bypassing the SDK library with direct API calls. Second, the provider's service derives caller identity from the kernel (Binder.getCallingUid()) and resolves the signing certificate via PackageManager, ensuring the attacker cannot inject false values even when invoking the SDK library. The attacker cannot spoof their UID, and their certificate hash will always reflect their own signing key, not the victim's.

**Table 2: Security properties of the three-layer defense.**

| Layer | Guarantees | Defeat requires |
| --- | --- | --- |
| L1: Binder UID | Caller process identity | Kernel exploit |
| L2: Certificate | Caller code authenticity | Signing key theft |
| L3: Server-side | Independent trust anchor | Server compromise |
| **Combined** | **All three properties** | **All three attacks** |

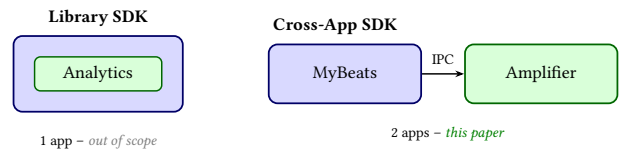

**Figure 4: Library SDKs embed within one app (out of scope). Cross-app SDKs require both apps; our defense replaces an insecure IPC path.**

Table 2 summarizes the security guarantees of each layer. Defeating this defense requires simultaneously forging a Linux UID (kernel exploit), possessing the victim's signing key, and compromising the backend registry—far beyond unprivileged application capabilities.

### 5.5 Deployment Considerations

The three-layer design decouples partner enrollment from SDK release cycles. When a new partner integrates the SDK, their (packageName, certHash) pair is registered server-side; no existing partner apps or SDK binaries require modification. This is critical for large-scale ecosystems where coordinating simultaneous updates across hundreds of partner apps is impractical. The server-side registry also provides an audit trail of authentication decisions, enabling forensic analysis if a compromise is later detected.

Importantly, our defense applies to SDKs requiring cross-app communication, where the SDK provider's application is already a prerequisite (e.g., content publishing, payment authorization, identity federation). The Bound Service approach does not introduce a new install requirement—it replaces an insecure IPC path with a secure one between apps that must coexist regardless. Library-only SDKs embedded entirely within partner APKs (analytics, crash reporting) face different trust constraints outside the scope of this work. Figure 4 illustrates this distinction.

## 6 Evaluation

We validated both the attack and defense through a proof-of-concept (PoC) implementation consisting of four Android applications deployed on a Pixel 7 running Android 14 (API 34).

Table 3: Attack success rate against SDK implementations.

| SDK Implementation | Attack Success | Defense Effective |
|---|---|---|
| VulnerableSDK (PI-based) | 50/50 (100%) | ✗ |
| SecureSDK (Binder-based) | 0/50 (0%) | ✓ |

## 6.1 PoC Components

(1) **PartnerApp** (com.poc.partner): A mock partner application that posts notifications containing PendingIntents.
(2) **VulnerableSDK** (com.poc.vulnerable.sdk): A mock SDK provider's service that authenticates callers via `PendingIntent.getCreatorPackage()`.
(3) **AttackerApp** (com.poc.attacker): A malicious app with NLS that harvests and replays PendingIntents.
(4) **SecureSDK** (com.poc.secure.sdk): A mock SDK provider's service using our three-layer defense with `Binder.getCallingUid()`.

## 6.2 Test Methodology

We conducted 50 attack attempts against each SDK implementation. For each attempt, PartnerApp posted a notification, AttackerApp harvested the PendingIntent via NLS, and AttackerApp invoked the SDK's publish API with the stolen credential and attacker-controlled content.

## 6.3 Results

Against VulnerableSDK, the attacker authenticated as PartnerApp in 100% of attempts. The service's `getCreatorPackage()` call returned `"com.poc.partner"` regardless of which app presented the token.

Against SecureSDK, 100% of attempts were rejected. `Binder.getCallingUid()` returned the attacker's own UID (10234), which resolved to `"com.poc.attacker"`—correctly identifying the actual caller rather than the PendingIntent's creator.

## 6.4 Performance Overhead

The caller verification itself (`Binder.getCallingUid()` and `PackageManager` lookup) adds negligible overhead (<1ms in our measurements). However, implementations should handle service binding asynchronously to avoid blocking the main thread, as synchronous cross-process IPC can contribute to ANR risk under high load. In our PoC, we used Kotlin coroutines to perform SDK calls off the main thread, eliminating any user-perceptible latency.

## 6.5 Limitations

Our PoC validates the attack and defense mechanisms but does not include a large-scale binary analysis quantifying how many production SDKs use the vulnerable pattern. Such a study—using tools like LibScout [15] or LibID [59] to identify `getCreatorPackage()` calls in authentication code paths—remains important future work. Scenario names (Amplifier, MyBeats) are fictitious. The vulnerable authentication pattern was identified through analysis of publicly documented SDK integration practices; the defense architecture has been validated in a realistic partner SDK environment.

## 7 Discussion

### 7.1 Implications for On-Device AI SDKs

On-device AI is transitioning from embedded libraries to cross-app platform services. Google's AICore, introduced in Android 14, exposes Gemini Nano as a shared system service that multiple apps access via Binder IPC [6, 7]. AICore uses system-level allowlists and Private Compute Core isolation for access control [7]—robust authentication, not PendingIntent.

However, if third-party vendors adopt similar architectures without equivalent rigor, the authentication question becomes critical. Commercial AI SDKs represent high-value targets: FaceTec maintains a $600,000 spoof bounty and holds 20+ patents on liveness detection [22]; IP litigation between FaceTec and Jumio underscores the economic stakes of these algorithms [29]. Biometric SDKs process regulated personal data (facial geometry, identity documents) under GDPR and KYC/AML requirements.

Crucially, on-device model extraction is a proven threat. Tramèr et al. established that ML models can be stolen via prediction APIs with high fidelity [53]. The DeMistify study extracted models from 82.7% of 1,511 apps tested, finding single models reused across 80+ applications [48]. SmartAppAttack successfully attacked 71.7% of 53 transfer-learning apps on Google Play [28]. Even encrypted models in Google Photos were extracted via runtime instrumentation [45]. A 2024 USENIX Security SoK concludes that on-device model extraction is "widespread, scalable, and under-defended" [38]. Recent defenses like SODA propose hardware-backed model protection, but these require TEE support not universally available [49].

The combination is concerning: if emerging cross-app AI services adopt weak IPC authentication—such as the `getCreatorPackage()` pattern we exploit—attackers could impersonate partners to extract proprietary models, consume inference quota, or access biometric data. Our Bound Service defense provides a template for how such services should authenticate callers.

### 7.2 Platform Mitigations and Their Limits

Android 15's new `RECEIVE_SENSITIVE_NOTIFICATIONS` permission and OTP masking [5] protect notification *content* (specifically OTP codes) but do not restrict access to the `PendingIntent` objects embedded in notifications. An NLS-enabled attacker on Android 15 can still extract PendingIntents from every notification—only the text content is masked.

Android 15 also extends the Restricted Settings framework to block *sideloaded* apps from obtaining NLS access, which partially mitigates our attack for sideloaded attackers. However, this restriction does not apply to apps distributed through the Play Store or compliant third-party stores—a malicious app that passes store review (as documented in prior work [30, 34]) retains full NLS access on Android 15. Since Play Store-distributed apps are the more capable threat vector (wider install base, no restricted-settings friction), this mitigation leaves the primary attack path unaffected.

OWASP's guidance for PendingIntent [39] addresses privilege escalation but not authentication bypass. A platform-level "presenter verification" API that attests the presenting process would close this gap but would require framework-level changes.

### 7.3 Broader Supply Chain Implications

Our findings connect to the broader mobile supply chain security crisis documented by OWASP and industry reports. The SourMint campaign demonstrated that a malicious advertising SDK could infiltrate 1,200+ iOS apps affecting 300 million monthly downloads before detection [46]. The X-Mode SDK collected and sold sensitive location data until platform bans and FTC enforcement action in 2024 [23]. The OneAudience and MobiBurn SDKs harvested social graph data, leading to litigation [36].

These incidents share a common pattern: SDK behavior that appears benign during review but enables unauthorized data collection or capability abuse at scale. Our provenance confusion attack adds another dimension to this threat: even SDKs with legitimate authentication *intent* may implement it insecurely, creating vulnerabilities that propagate across their entire partner ecosystem. OWASP's M2 ranking reflects this reality: third-party SDKs are not merely dependencies but active attack surfaces that require dedicated security analysis [40].

### 7.4 Ethical Considerations

The attack described in this paper was demonstrated exclusively against synthetic applications that we developed for this purpose. No production SDK or partner application was targeted during our evaluation. The vulnerable authentication pattern was identified through code review of publicly documented Android APIs and SDK integration guides, not through reverse engineering of proprietary binaries. We use fictitious scenario names throughout the paper.

*Responsible Disclosure.* Since the vulnerability resides in an application-layer design pattern—not in the Android platform itself—there is no single vendor to notify. Android's official documentation explicitly warns against using `getCreatorPackage()` for authentication, noting that attackers can obtain PendingIntents via `NotificationListenerService` and that this can lead to authentication bypass [11]. We provide a courtesy notification to Meta, whose one-tap authentication on WhatsApp uses this pattern; Meta is independently deprecating the PendingIntent-based handshake on April 15, 2026 in favor of the OTP Android SDK [35].

### 7.5 Evidence of Real-World Pattern Usage

The vulnerable pattern exhibits a documentation-driven propagation model: platform SDK documentation and developer Q&A actively teach `getCreatorPackage()` as a sender verification mechanism. This guidance propagates through SDK ecosystems, creating systemic exposure.

*Platform SDK documentation.* WhatsApp's one-tap autofill authentication uses `getCreatorPackage()` for partner app verification. Meta's documentation states that "the match is determined through the `getCreatorPackage` method called on the PendingIntent object provided by your application" [35]. Sample code in the `OtpErrorReceiver` explicitly checks whether `getCreatorPackage()` returns `"com.whatsapp"` or `"com.whatsapp.w4b"` before trusting callback data. This pattern propagates through business messaging SDKs: Alibaba Cloud's ChatApp documentation [3], Webex Connect [55], and YCloud all instruct developers to implement identical `getCreatorPackage()`-based verification. Meta is deprecating the PendingIntent-based handshake in April 2026—an implicit acknowledgment of the pattern's security limitations.

*Open-source implementations.* Chromium's `IntentHandler` checks `getCreatorPackage()` for token validation [18]. The Termux plugin architecture uses it to verify plugins receive PendingIntents only from the official Termux app [50]. Android's own framework uses the pattern: AOSP's `MediaButtonReceiverHolder` filters media button receivers to only accept components whose package matches `getCreatorPackage()`—a trust constraint that could be bypassed via stolen PendingIntents [14].

*Developer guidance.* Stack Overflow's highest-voted answer for "how to get the sender of an Intent" recommends `getCreatorPackage()` as the solution, stating that the creator identity "cannot be forged by the app" [1]. This advice, viewed by thousands of developers, conflates unforgeability of creator identity with unforgeability of presenter identity—precisely the provenance confusion we exploit.

*Academic measurement.* PITracker [56] analyzed 10,000 third-party apps, detecting 2,939 PendingIntent-related threats with 11 confirmed exploits. The Stickytent study [4] of 23,922 apps found 17% of PendingIntent vulnerabilities lead to unauthorized access. Notably, these studies report vulnerability counts but do not measure `getCreatorPackage()` usage specifically—a targeted static analysis quantifying authentication-context usage remains future work.

*Vendor response.* Samsung has disclosed four PendingIntent hijacking CVEs since 2021 (CVE-2021-25355, CVE-2021-25364, CVE-2022-23434, CVE-2023-21466) [44]. Android's security documentation explicitly warns against using `getCreatorPackage()` for authentication, noting that attackers can obtain PendingIntents via NotificationListenerService [11]. A 2021 AOSP commit enforces package visibility restrictions on `getCreatorPackage()`, treating the API as security-sensitive [13].

### 7.6 Limitations

Our attack requires that the target partner app posts notifications containing a `PendingIntent` (near-universal for Android apps); a partner that never posts notifications would not be vulnerable through this specific vector, though other PendingIntent leakage paths exist [56]. Our defense assumes the SDK author controls a Bound Service component; SDKs distributed as pure library code without a Service component would require architectural changes.

## 8 Related Work

*PendingIntent Security.* PITracker [56] formalized a PendingIntent vulnerability taxonomy, identifying 72 CVEs and 2,939 vulnerable objects across AOSP, and confirming that NLS can extract PendingIntents from notifications. PendingMutent [2] analyzed 180,606 apps and proposed a capability-based authorization framework. He et al. [26] demonstrated universal PendingIntent exploitation for privilege escalation. These works focus on privilege escalation and data leakage; we are the first to formalize provenance confusion as an authentication-bypass vector in SDK ecosystems and to present a validated defense.

*NotificationListenerService Abuse.* NLRadar [32] is the most comprehensive study of NLS abuse, combining static analysis with LLMs to analyze apps at scale. Knock-Knock [17] first demonstrated notification data leakage risks. MITRE ATT&CK documents NLS abuse as technique T1517 [37]. Active malware campaigns including Tria Stealer [31], Aberebot [20], S.O.V.A. [19], and OtpSteal [52] have weaponized NLS for credential theft. These works focus on data exfiltration rather than authentication bypass.

*Android IPC and Component Security.* Felt et al. [16] analyzed Android's IPC authentication model and established that Binder provides kernel-enforced caller identity. BinderCracker [25] fuzzed 170+ system services and confirmed that `Binder.getCallingUid()` is unspoofable from unprivileged context. ComponentSecurity [27] found 8.8% of 13,835 popular apps vulnerable to cross-layer component attacks. VenomAttack [33] demonstrated automated activity hijacking. Our work leverages the kernel-level guarantees validated by these studies to construct a positive defense architecture.

*SDK Supply Chain Security.* OWASP's 2024 Mobile Top 10 ranks Inadequate Supply Chain Security as M2 [40], explicitly identifying third-party SDKs as primary threat vectors. Sonatype's 2024 report documents 704,000+ malicious packages with 156% YoY growth [47]. Zhan et al. [57] systematically reviewed 74 papers on Android third-party libraries. ATVHunter [58] found 8.7% of apps contain vulnerable library versions. The Goldoson [34], Necro Trojan [30], and SourMint [46] incidents demonstrate that SDK vulnerabilities propagate at ecosystem scale. X-Mode's FTC ban [23] and the OneAudience/MobiBurn litigation [36] show regulatory consequences of SDK abuse. Our contribution is orthogonal: we address authentication at the IPC boundary between partner apps and SDK services, a trust relationship these works do not examine.

## 9 Conclusion

We have demonstrated that `PendingIntent`-based SDK authentication suffers from a fundamental provenance confusion vulnerability: the API conflates token *creation* with token *presentation*, enabling attackers with notification access to impersonate legitimate partners. This vulnerability is particularly concerning given OWASP's 2024 ranking of Inadequate Supply Chain Security as the second most critical mobile risk and the documented history of SDK-driven compromises affecting hundreds of millions of users.

Our systematic analysis of eight IPC authentication mechanisms reveals that only Bound Service IPC with kernel-level Binder UID verification satisfies all security properties required for SDK partner authentication: kernel-backed identity, unforgeability from application context, replay resistance, scalability without manifest changes, and bidirectional communication support.

The three-layer defense we present—combining Binder UID verification, certificate hash validation, and server-side partner registry—provides defense-in-depth that an attacker cannot defeat without simultaneously compromising the Linux kernel, stealing signing keys, and breaching backend infrastructure. Importantly, this architecture supports the operational realities of SDK ecosystems: new partners are onboarded server-side without requiring SDK updates or coordinated partner app releases.

As on-device AI SDKs proliferate—distributing proprietary model weights for local inference on vision, language, and biometric tasks—the stakes of SDK authentication failures will only increase. We hope this work contributes to more secure SDK architectures and encourages platform vendors to consider native support for presenter verification in future Android releases.